\documentclass[aps,prl,preprint,groupedaddress,superscriptaddress]{revtex4-1}
\usepackage{graphicx}
\usepackage{color}

\begin{document}
\title{Enabling single-mode behavior over large areas \\ with photonic Dirac cones}
\author{J. Bravo-Abad}
\affiliation{Departamento de Fisica Teorica de la Materia Condensada, Universidad Autonoma de Madrid, 28049 Madrid, Spain}
\author{J. D. Joannopoulos}
\affiliation{Department of Physics, Massachusetts Institute of Technology, Cambridge, MA 02139  USA}
\author{M. Solja\v{c}i\'{c}}
\affiliation{Department of Physics, Massachusetts Institute of Technology, Cambridge, MA 02139  USA}

\begin{abstract}
Many of graphene's unique electronic properties  emerge from its Dirac-like electronic energy spectrum~\cite{Novoselov04,Zhang05,Novoselov05,Berger06,Geim07,CastroNeto09,Geim09}. Similarly, it is expected that a nanophotonic system featuring Dirac dispersion will open a path to a number of important research avenues~\cite{Haldane08,Sepkhanov07,Dood10,Bahat-Treidel10,Zhang08,Huang11}. To date, however, all proposed realizations of a photonic analog of graphene lack fully omnidirectional out-of-plane light confinement, which has prevented creating truly realistic implementations of this class of systems. Here we report on a novel route to achieve all-dielectric three-dimensional photonic materials featuring Dirac-like dispersion in a quasi-two-dimensional system. We further discuss how this finding could enable a dramatic enhancement of the spontaneous emission coupling efficiency (the $\beta$-factor) over large areas, defying the common wisdom that the $\beta$-factor degrades rapidly as the size of the system increases. These results might enable general new classes of large-area ultralow-threshold lasers, single-photon sources, quantum information processing devices and energy harvesting systems.
 \end{abstract}
 
 \maketitle

Since its isolation from bulk graphite in 2004~\cite{Novoselov04}, graphene --a one-atom-thick sheet of carbon-- has attracted an ever increasing  amount of interest~\cite{Zhang05,Novoselov05,Berger06,Geim07,CastroNeto09,Geim09}; nowadays, the study of the electronic properties of this two-dimensional (2D) material has become one of the most active areas of condensed matter physics. This general endeavor has also stimulated new directions in related research fields, especially those originally inspired by the physics of electronic transport in crystalline solids. Of particular relevance in this context are photonic materials whose dielectric constant is periodically structured at the subwavelength scale, the so-called \emph{photonic crystals} (PhCs)~\cite{JJbook}. By exploiting the analogy between the propagation of electrons in graphene and the propagation of photons in suitably designed 2D PhCs, phenomena such as directional optical waveguiding~\cite{Haldane08},  pseudodiffusive transport of light~\cite{Sepkhanov07,Dood10}, Klein tunneling~\cite{Bahat-Treidel10},  and the observation of the \emph{Zitterbewegung} of photons~\cite{Zhang08}  have been recently proposed. The existence of Dirac points at the center of the Brillouin zone induced by accidental degeneracy in square lattice 2D PhCs, has also been discussed recently~\cite{Huang11}. However, all these systems share the common fundamental drawback that they lack fully omnidirectional out-of-plane light confinement, which has so far prevented the creation of a truly realistic implementation of a photonic counterpart of graphene.

In this Letter, we propose a feasible approach to achieve simultaneously quasi-two-dimensional light propagation and Dirac cone dispersion in an all-dielectric 3D photonic material particularly suitable for optical device integration. We show how the unique light confining properties of a proper choice of 3D layered PhCs enables creating extended planar defect modes whose dispersion relation exhibits isolated Dirac points inside a complete 3D photonic band-gap. 
In the limit in which the emitter frequency virtually coincides with the Dirac point frequency (i.e., the frequency of the Dirac cone vertex) the number of photonic states available to the emitter approaches one, even if the system features a macroscopic area. Thereby, the photonic materials presented in this Letter enable for the first time the implementation of structures much larger than the wavelength, which nevertheless have $\beta$ factors close to unity. Due to the crucial role played by the $\beta$-factor in various areas of physics (from optoelectronics to quantum computation or energy harvesting), we believe that these results hold a great promise for the development of novel types of nanodevices.

\begin{figure}[t]
\includegraphics[width=17cm]{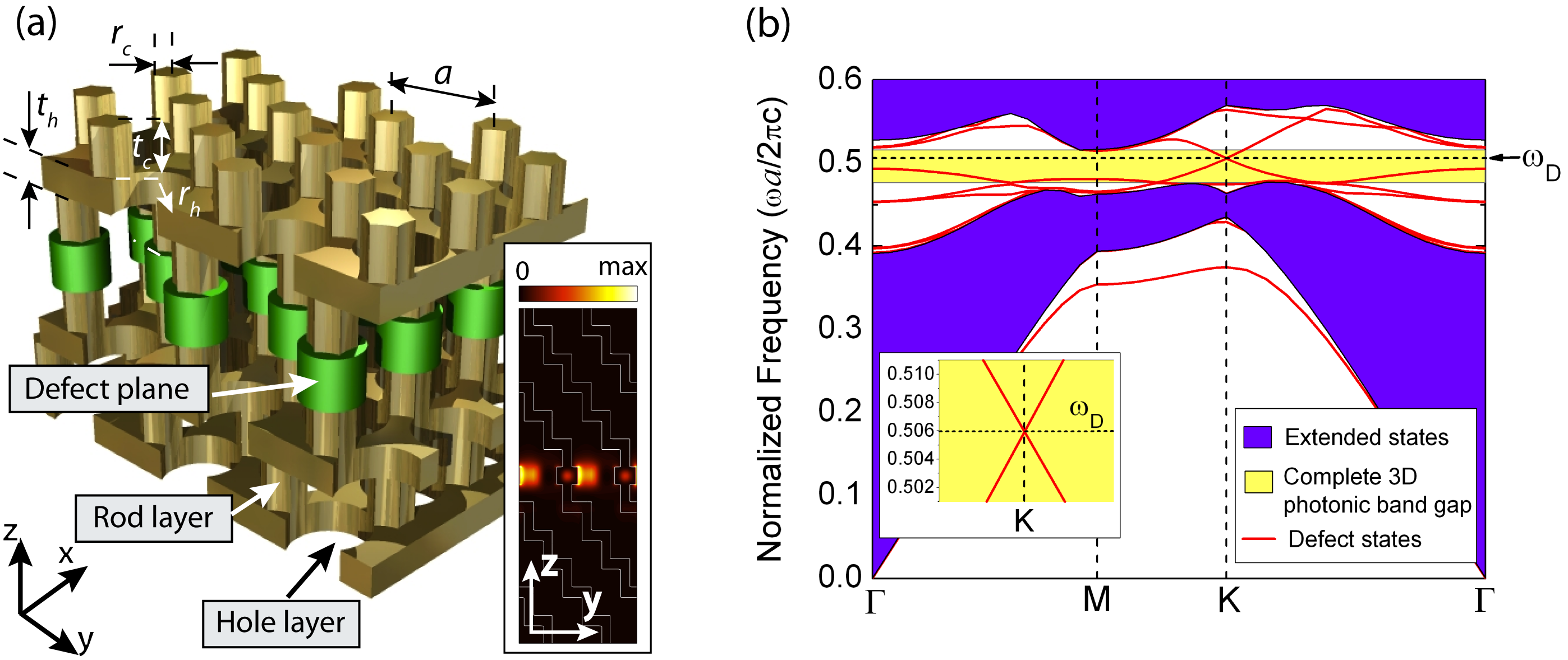}
\caption{ {\bf An isolated photonic Dirac point in a 3D photonic crystal}. Panel (a) displays a sketch of the considered structure. The basic structure consists of face-centered cubic (fcc) lattice of overlapping air cylinders embedded in a high dielectric background (yellow regions). This structure can also be seen as an alternating stack of rod and hole layers (see labels in the figure). A planar defect is introduced in the structure by removing a hole layer and replacing it by a triangular array of finite-height dielectric cylinders (green cylinders). The geometrical parameters used to define the structure are also displayed. The inset renders the electric-field intensity corresponding to a guided mode in the defect plane at the Dirac frequency. Panel (b) displays the dispersion diagram corresponding to the structure shown in Panel (a), projected over the First Brillouin zone of the in-plane triangular lattice characterizing the hole and rod layers  of that structure. The inset of Panel (b) shows an enlarged view of the dispersion diagram near the Dirac point.}
\end{figure}

A schematics of the considered photonic material is rendered in Fig.~1a. We start by considering the electromagnetic properties of the PhC before the defect plane is introduced; it is a face-centered cubic (fcc) 3D PhC of air (or low-index) cylinders embedded in a dielectric background and oriented along the (1 1 1) direction~\cite{Johnson00,Qi04}. This peculiar class of layered PhCs can actually be viewed as an alternating stack of two different types of layers. One of the layers has the form of a triangular lattice of finite-height dielectric rods in air (labeled as \emph{rod layer} in Fig. 1a), whereas the other layer can be described as a triangular lattice of air holes milled in a dielectric slab (labeled as \emph {hole layer} in Fig.~1a). This PhC features two important characteristics. First, each of the two types of layers displays a highly symmetric cross section that mimics a canonical 2D photonic-crystal structure: one is a periodic array of air holes in a dielectric dielectric slab, and the other is a periodic array of hexagonal-like rods in air. Second, although neither of the layers displays a complete (omnidirectional) photonic band gap by itself, when the layers are periodically stacked as shown in Fig.~1a, a large complete photonic band-gap can be obtained, using practical values of the refractive index contrast ~\cite{Johnson00,Qi04}. 

Next, consider creating an extended planar electromagnetic defect mode in the 3D PhC described above. In order to that, we introduce into the structure a single defect layer that perturbs the original periodic sequence rod-layer/hole-layer/rod-layer along the (1 1 1) direction. In particular, we remove a hole-layer of the structure and replace it by a triangular lattice of finite-height dielectric rods with circular cross section (green cylinders in Fig.~1a), whose radius, height and dielectric constant are given by $r_d$, $h_d$ and $\epsilon_d$, respectively.

It is known~\cite{Povinelli01,Qi04} that  the introduction of line defects into layered 3D PhCs of the type described above enables the implementation of localized electromagnetic states whose dispersion relation, field profiles, and polarization are in a close correspondence with those associated with the corresponding 2D PhC geometries.  Similarly, one would expect that the spectral properties of the planar extended states localized in the defect layer rendered in Fig.~1a should, to some extent, inherit properties associated with the Bloch states of the bona-fide 2D counterpart of this defect layer (i.e., those that are present in a triangular array of infinitely long high-index rods). On the other hand, it has been demonstrated that the intrinsic symmetry properties of 2D PhCs based on a triangular lattice can induce the presence of Dirac points near high-symmetry points of the band structure~\cite{Haldane08,Sepkhanov07,Dood10}. Thus, this line of reasoning  suggests the feasibility of creating a single 3D physical system simultaneously featuring quasi-two-dimensional light propagation and Dirac cone dispersion. In addition to these two features, it is also crucial that the mentioned Dirac cone dispersion is isolated within a given frequency bandwidth, or, equivalently, that the Dirac cone is fully separated from the rest of the bands present in the band structure. Only by combining all three of these features in the same system, it is possible to fully exploit the analogy between electronic and photonic graphene. We emphasize that, although, in principle, it is not challenging to obtain each of these three characteristics separately in a photonic structure, achieving all three of them concurrently in the same 3D physical system is not straightforward.

To explore to which extent these ideas can be implemented in a realistic system, we have carried out a systematic numerical analysis of the evolution of the corresponding band structures as a function of the geometrical parameters and the dielectric constants, defining both the underlying 3D PhC, as well the defect layer. The calculations were performed by means of the plane-wave expansion method to Maxwell's equations~\cite{Johnson01} using a supercell large enough in the (1 1 1) direction so the properties of an isolated defect plane in an infinite 3D PhC are accurately reproduced. Our calculations show that, for the optimal structure, the air holes within the hole layer and the {\emph {equivalent-cylinders}}~\cite{Povinelli01} in the rod layer are $r_h=0.41 a$ and $r_c=0.18 a$ respectively  ($a$ is the lattice constant of the in-plane triangular lattice defined within each layer; see Fig.~1a). The thicknesses of the hole and rod layers are $t_h =0.32 a$   and $t_c=0.50 a$, respectively, whereas the refractive index of the high-dielectric material is assumed to be $n=2.5$. The low-refractive index of the structure is taken to be air. On the other hand, the defect layer of this optimal system features $n_d=3.1$, $r_d=0.32 a$, and $h_d=t_c$. In addition, from this numerical study we have also obtained that  the optimal configuration is that in which the cylindrical rods of the defect layer are aligned with the rods forming the two rod-layers located right above and below the defect plane (see Fig.~1a).

Figure~1b summarizes the dispersion diagram obtained for this structure. Shaded violet areas in this figure show the projected band structure for the perfectly periodic 3D PhC (i.e., without the defect layer). In this case, the dispersion diagram was obtained by plotting the frequencies $\omega$ of the extended bulk states of the system as a function of the in-plane wavevector ${\bf k}_{\parallel}$ in the irreducible Brillouin zone of the underlying in-plane 2D triangular lattice. The considered system exhibits a large 3D complete photonic band gap (shaded yellow area), centered at frequency $\omega=0.497 (2 \pi c/a)$ ($c$ is the light velocity in vacuum) and featuring a gap-midgap ratio of  approximately 8\%. The dispersion relations of the guided modes of the defect plane are also rendered in Fig.~1b (red lines). As can be seen, these defect bands indeed display a Dirac point at $\omega_D=0.506 (2 \pi c/a)$ (see  Fig.~1b and the inset of Fig.~1a), fully lying within the omnidirectional photonic-band gap of the periodic system (yellow area in Fig.~1b). Importantly, this Dirac point is completely isolated from all of the rest of the frequencies of the band structure of the system within a bandwith $\Delta\omega=0.026 (2 \pi c/a)$. Finally, the strong out-of-plane photonic-band gap confinement of the electromagnetic fields at  $\omega=\omega_D$ is clearly observed in the inset of Fig.~1a, which displays the corresponding cross-section along the $xy$ plane of the electric-field intensity distribution. We believe that these results represent the first time that a complete photonic analog of graphene is implemented in a realistic 3D physical system. We emphasize that the obtained results are scalable to many different frequency regimes, and therefore, could be used to enhance performance of different classes of active optical devices.

We now turn to demonstrate how the proposed class of photonic systems can enable an unprecedented control of light-matter interaction over large areas. In order to do that, we apply the commonly employed procedure to probe the properties of light-matter interaction in a complex  electromagnetic environment: we study the radiation process of a point quantum emitter embedded in the system under analysis~\cite{Novotnybook,Yariv}. The emitter is modeled as a two-level system, characterized by a transition frequency $\omega_s$, an emission bandwidth (i.e., a transition bandwidth) of $\Delta \omega_s$, and a dipolar transition moment ${\bf d}=d \: \hat{{\bf d}}$. In particular, we analyze the spontaneous emission coupling efficiency, the so-called $\beta$-factor. In a given multimode photonic system, the $\beta$-factor quantifies the portion of all spontaneously emitted photons that couple into a certain {\emph {targeted}} mode~\cite{Coldrenbook}. This physical magnitude is of a paramount importance in modern optoelectronics and quantum information processing since the performance of an active nanophotonic device, or a single photon emitter, can often be greatly enhanced by increasing the value of $\beta$~\cite{Coldrenbook,Yokoyama92, Baba91} (recent examples include ultralow-threshold lasers~\cite{Painter99,Vuckovic06,Vuckovic11} and single photon sources~\cite{Santori02,Chang06} based on PhC cavities).

The dependence of the $\beta$-factor on the particular electromagnetic environment in which the considered emitter is embedded can be elucidated by examining its link with the corresponding photonic local density of states (LDOS), $\rho({\bf r}, {\hat {\bf d}},\omega)$. For a non-dissipative system, the LDOS can be written as~\cite{Novotnybook}
\begin{equation}
\rho({\bf r}, {\hat {\bf d}},\omega)=\sum_{\nu} \epsilon({\bf r}) \: |{\bf E}_{\nu}({\bf r}) \cdot {\hat{\bf d}}|^2 \: \delta(\omega-\omega_{\nu}) 
\end{equation}
where the index $\nu$ labels the different source-free normal solutions to Maxwell's equations obtained for the considered photonic structure; ${\bf E}_{\nu}({\bf r})$  and $\omega_{\nu}$ are their corresponding $E$-field profile and frequency, respectively. $\epsilon({\bf r})$ stands for the dielectric constant distribution.

On the other hand, from Fermi's golden rule~\cite{Purcell,Novotnybook,Yariv}, one finds that, in 3D, the spontaneous emission rate of the considered quantum emitter $\Gamma$ is proportional to the LDOS accessible to the emitter  \mbox{$\Gamma= (\pi |{\bf d}|^2 \omega_s/ 3 \hbar \epsilon_0) \: \rho({\bf r_s},{\hat {\bf d}},\omega_s)$}. Thus, assuming the targeted mode to be a normal mode of the system, ${\bf E}_t({\bf r})$, of frequency $\omega_t$, one can write the $\beta$ factor as (see Suppl. Information),
\begin{equation}
\beta= \frac{\omega_t \: g(\omega_t) \: \epsilon({\bf r}) \: |{\bf E}_t({\bf r}) \cdot {\hat{\bf d}}|^2} {\int d \omega \: g(\omega) \: \omega \: \rho({\bf r_s},{\hat {\bf d}},\omega)}
\end{equation}
where $g(\omega)$ is the lineshape of the transition, centered at $\omega_s$ and characterized by full-width-half-maximum (FWHM) of $\Delta\omega_s$. Note that the factor $g(\omega_t)$ in the numerator accounts for the fact that  $\beta$ decreases as the emission frequency is detuned from $\omega_t$. Hereafter $\omega_t=\omega_s$ is assumed.

Equation~(2) clearly shows that the $\beta$ factor can be enhanced by introducing a physical mechanism that minimizes  the density of photonic states lying within the transition linewidth. In fact, the large values for $\beta$ reached in subwavelength volume photonic resonators~\cite{Painter99,Vuckovic06,Santori02,Chang06} can be viewed as a particular instance of this physical picture. Such nanoresonators are designed to have a volume small enough so that only one resonant mode lies within the transition linewidth. This makes them to effectively act as single-mode structures, which as deduced from Eq.~(1) leads to values $\beta \approx 1$ (provided that the coupling with the radiation modes existing outside the resonator is negligible). Although very relevant in a number of contexts, this cavity-based approach does not admit a straightforward extension for large-area control of $\beta$. In the rest of this Letter, we show how the general class of isolated Dirac points discussed above do enable such large-area control. 

\begin{figure}[t]
\includegraphics[width=14cm]{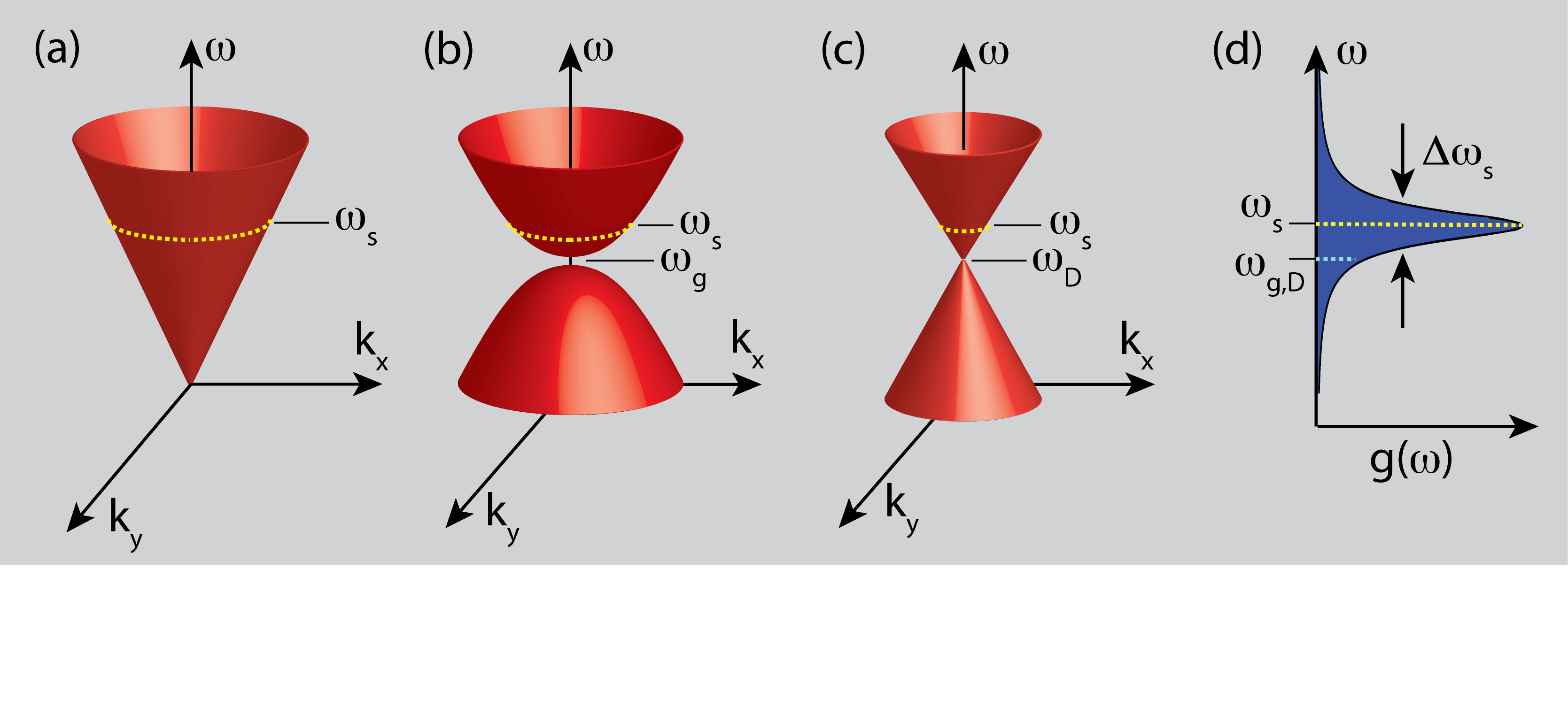}
\caption{ {\bf Dispersion relations of three representative photonic materials and typical spectrum of a quantum emitter}. Panel (a) and (b) sketch, respectively, the dispersion relations of a homogeneous material and a photonic-crystal exhibiting a photonic band gap near the peak of the emission $\omega_s$. Panel (c): same as (b) but for the case in which the structure exhibits a Dirac point near $\omega_s$. Panel (d) renders the lineshape of the emission spectrum $g(\omega)$ in the form of a Lorentzian centered at $\omega_s$ and featuring a full-width at half-maximum (FWHM) of $\Delta\omega_s$. $\omega_g$ and $\omega_D$ denote the central frequency of the band-gap and the Dirac point frequency, respectively.}
\end{figure}

To gain physical insight into the effect of the Dirac cone dispersion on the $\beta$ factor, we evaluate the magnitude of $\beta$ in the following three cases (all of them 3D): first, the case of a homogeneous material; second, the case in which the dielectric material is periodically structured so that $\omega_s$ lies in the vicinity of the lower edge of a 3D photonic-band gap; and, third, the case of the structure displayed in Fig.~1, i.e., a system exhibiting simultaneously quasi-two dimensional light propagation and an isolated Dirac point near $\omega_s$. Figures~2a,~2b, and~2c illustrate each of these dispersion relations (for simplicity in the visualization, in Figs.~2a and~2b, only the 2D counterparts of the corresponding cases are rendered). Importantly, in all three cases we also assume that the EM field in the system is confined in a finite volume $V$, such that $V >> \lambda^3$ (the dependence on volume is addressed below).

The homogeneous case is characterized by the following dispersion relation: $\omega(k)=ck/n$ (where $n$ is the refractive index and $k=|{\bf k}|$, with ${\bf k}=({\bf k}_\parallel,k_z)$). Then, making use of the isotropy of the medium, one can derive  the following simple expression for the $\beta$ factor: $\beta=(1/V) \:\omega_s g(\omega_s)/  F_h(\omega)$, where $F_h(\omega) \equiv \int d\omega\:  g(\omega) \: \omega \: \tilde{\rho}_h(\omega)$, with $\tilde{\rho}_h(\omega)=(1/2\pi)(n/c)^3\omega^2$. Similarly, for the case of the photonic band-gap, taking  \mbox{$\omega(k)=\omega_g-A_g k^2$} ($\omega_g$ being the center of the gap and, see Fig.~1b, and $A_g$ is a constant which has to be determined from calculations of the band structure; physically, $A_g$ defines the curvature of the dispersion relation close to $\omega_g$), one finds that the magnitude of $\beta$ can be calculated using the same expression given above for the homogeneous case but replacing $F_h(\omega)$ by $F_g(\omega) \equiv \int d\omega\:  g(\omega) \: \omega \: \tilde{\rho}_g(\omega)$, with \mbox{$\tilde{\rho}_g(\omega)=(4\pi^2 A_g^{3/2})^{-1} (\omega_g-\omega)^{1/2}$}. 


For the Dirac case, the calculation of $\beta$ is more involved that in the previous two cases. First, we take into account that an excited dipole embedded in the defect layer displayed in Fig. 1a only decays via the guided modes confined within the layer; the decay into any other modes surrounding the layer (e.g., bulk Bloch modes) is prevented by an omnidirectional photonic band gap. Then, for small enough values of $h_d$ (so only the fundamental mode guided mode in the $z$-direction is excited), the quasi-2D light propagation inside the defect layer can be described by the dispersion relation corresponding to in-plane Bloch states (i.e., states with $k_z=0$): $\omega({\bf k}_\parallel)=\omega_D \pm A_D |{\bf k}_{\parallel}-{\bf k}_{\parallel,0}|$ (see details in Suppl. Information). In this expression, $A_D$ is a constant that can be obtained from band-structure calculations and ${\bf k}_{\parallel,0}$ defines the coordinates in $k$-space of the Dirac cone vertex, whereas the plus and minus signs correspond to $\omega > \omega_D$ and $\omega < \omega_D$, respectively. (Note that, physically, $A_D$ corresponds to the slope of the Dirac cone). Thus, one finds that the total density of states accessible to the emitter as 
\mbox{$\tilde{\rho}_D(\omega)=1/(2 \pi A_D^2)|\omega-\omega_D|$}. This, in turn, yields the following expression for the SE coupling efficiency 
$\beta=(1/A) \omega_s g(\omega_s)/F_D(\omega)$, with  $F_D(\omega)=\int d\omega g(\omega)  \omega  \tilde{\rho}_D(\omega)$. Here $A$ is the transversal area of the defect layer (i.e., the total volume of the defect layer is $V=A \times h_d$). 

We now quantify the values of $\beta$ for each of the above cases using realistic parameter values. To allow for a direct comparison among the three considered class of systems, we introduce a renormalized spontaneous emission coupling efficiency $\eta$. In the homogeneous and band-edge cases we define this magnitude as $\eta_{h,g}=\beta_{h,g} \times V/a^3$, whereas for the Dirac case we define $\eta_D=\beta_D\times S/a^2$ . This normalization allows us, on one hand, to remove from the discussion the obvious dependence of $\beta$ on the size of the system, and thus, focus exclusively on the photonic properties. On the other hand, it also removes the geometrical factor $V/A$ that enhances the $\beta$ factor in the Dirac case which respect to the other two cases. This factor stems from the electromagnetic confinement in the $z-$direction of the guided modes in the Dirac structure, and therefore cannot be ascribed to the Dirac spectrum. Furthermore, in our calculations we assume that the transition lineshape is described by a Lorenztian centered at $\omega_s=2.1\times 10^{15}$~s$^{-1}$  (i.e., an emission wavelength of 900 nm) and featuring a relative FWHM $\Delta\omega_s/\omega_s=10^{-4}$; values more than one magnitude smaller for  $\Delta\omega_s/\omega_s$ can be reached using, for instance, quantum dots at low temperatures. In the homogeneous case a refractive index $n=n_d$ is chosen, whereas for the band-edge and Dirac cases, the values of the dispersion relation parameters $A_g$ and $A_D$ are taken from band structure calculations: $A_g=1.2 \times c a/(2 \pi)$  and $A_D=0.3 \times 1/c $ ($a$ is lattice constant defined in Fig.~1a, which for operation at the considered emission wavelength, takes a value of 450 nm).  

Thus, using these parameters, from the numerical evaluation of the expression for $\beta$ given above, we obtain $\eta_h=68.2$, $\eta_g=4.0 \times 10^5$ and $\eta_D=3.6 \times 10^6$. As readily deduced from these values, the Dirac dispersion introduces an enhancement factor of about four orders of magnitude with respect to the homogeneous case and about one order of magnitude with respect to the band-edge case. This is an important result, since, as we show below, enables reaching values of $\beta \approx 1$ over macroscopic areas. Physically, the origin of this dramatic increase of $\eta$ (and consequently of $\beta$) can be understood in terms of the rapid decrease of the number of photonic states available to the emitter as its emission frequency approaches the frequency of the Dirac point. In particular, in contrast to the homogeneous and band-edge cases, in the Dirac case when $\omega_s \to \omega_D$ (i.e., as emission frequency approaches the Dirac vertex frequency), the number of modes accessible to the emitter approaches unity, making the whole structure to effectively behave as a single-mode system, even if it features a large area. Note that the LDOS is strictly zero at $\omega_g$ and $\omega_D$ in the band-edge and Dirac cases, respectively. Therefore, in the calculations for each case, we have assumed $\omega_s$ to be slightly detuned (by a quantity much smaller than $\omega_s$ and $\Delta\omega_s$) from $\omega_g$ and $\omega_D$.

\begin{figure}[t]
\includegraphics[width=12cm]{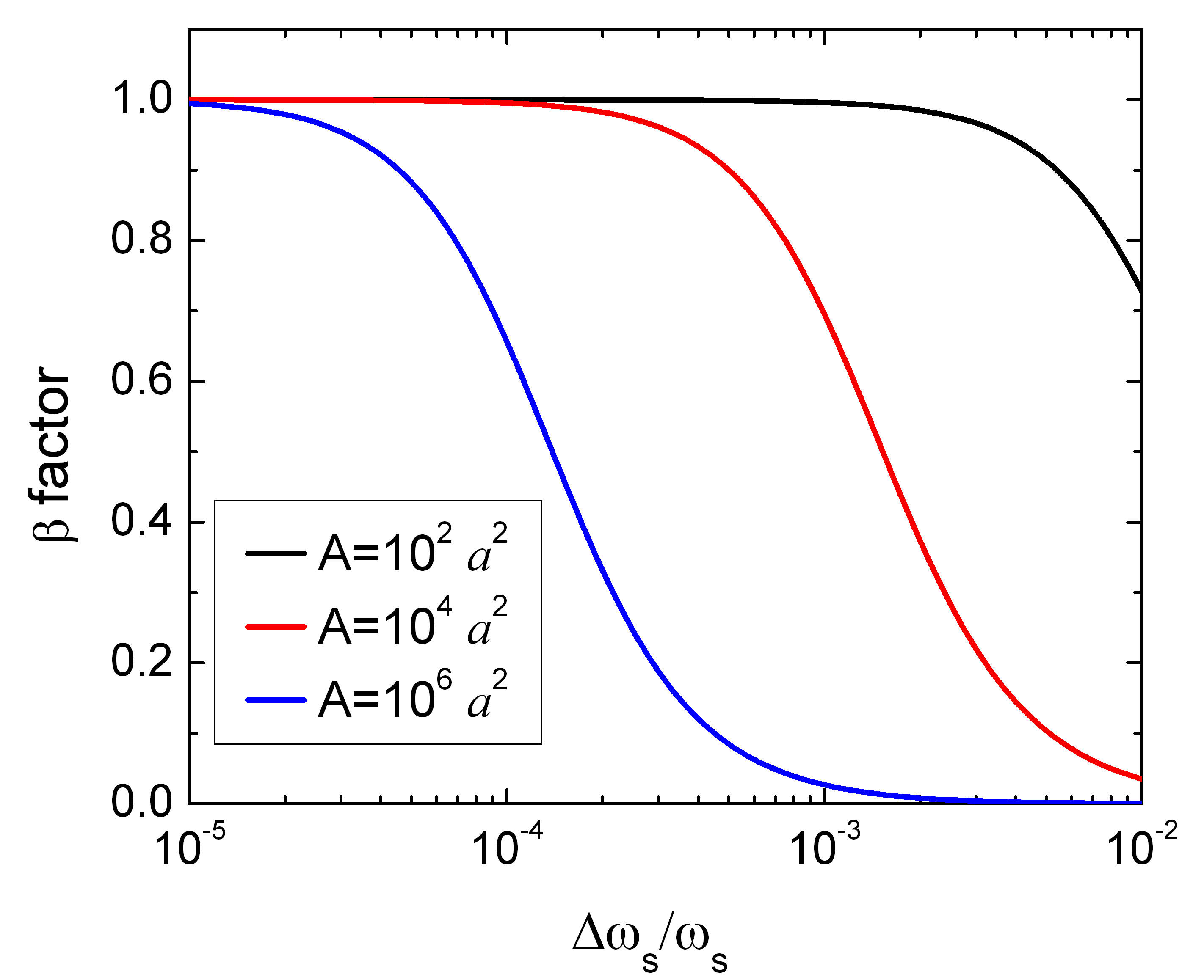}
\caption{{\bf Finite size effects on the $\beta$-factor.} Dependence of the $\beta$-factor on the normalized emission linewidth $\Delta \omega_s/ \omega_s$, as computed for several sizes of the transverse area $A$ of the system, with $a$ being the periodicity of the in-plane triangular photonic crystal (see definition in Fig.~1a).}
\end{figure}

We now focus on analyzing the dependence on size of the enhancement of $\beta$ enabled by isolated photonic Dirac cones. The physical origin of this dependence stems from the fact that for values of the area $A$ such that  $A>>\lambda^2$ cannot be safely assumed, it is necessary to account for the discreteness of the eigenmodes in the system: the only states that will be allowed to exist are those characterized by a wavevector ${\bf k}=(k_x,k_y)$  whose value coincides with one of the nodes of the rectangular grid defined by the discrete set of values $\{2\pi n_x/L,2\pi n_y/L\}$ (with $n_x$ and $n_y$ being arbitrary integers, and where the system is assumed to be square shaped with side length $L$, i.e., $A=L^2$)~\cite{kittelbook}. Therefore, the influence of these finite-size effects on $\beta$ can be computed by using the discrete version of Eq.~(2) (see Suppl. Information).

Figure~3 shows the computed results for $\beta$ as a function of the normalized emission bandwidth $\Delta\omega_s/\omega_s$ for several values of the lateral size of the system, ranging from $L/a=10$ to $L/a=10^3$. As observed, in the considered cases, $\beta$ tends to 1 when $\Delta\omega_s/\omega_s$ approaches the lower limit of the interval displayed in the figure ($\Delta\omega_s/\omega_s=10^{-5}$). This is due to the fact that for all the considered system sizes, a linewidth $\Delta\omega_s/\omega_s \lesssim 10^{-5}$ is smaller than the frequency interval between the adjacent modes, and therefore, the structure is actually acting as a single mode system (much in the same way as occurs in large-$\beta$ photonic nanocavities). As $\Delta\omega_s/\omega_s$ is increased, a growing number of modes \emph{enter} into the interval where $g(\omega)$ is not negligible and, therefore, the value of $\beta$ starts decreasing. Since the frequency interval between adjacent modes is smaller for larger values of $A$, the decrease of the $\beta$ factor with $\Delta\omega_s/\omega_s$ starts sooner for larger values of $A$.

In conclusion, we have demonstrated for the first time that a 3D photonic material can exhibit isolated Dirac cones in its dispersion relation. The proposed approach is readily accessible experimentally, both for fabrication and measurement. In addition, we have shown how the proposed class of systems enables achieving unprecedentedly large values of $\beta$-factor over large areas. We believe that these results could open new avenues in a variety of different fields. Thus, the large $\beta$ factors demonstrated in this work may lead to the realization of important applications such as  low-threshold large-area lasers, enhanced single-photon sources and novel efficient platforms for solar energy harvesting.

\begin{acknowledgments}
We acknowledge helpful discussions with Dr. L. Lu and S. L. Chua. JBA was supported in part by the MRSEC Program of the National Science Foundation under award number DMR-0819762 and in part by the Ramon-y-Cajal program of the Spanish MICINN under grant no. RyC-2009-05489. MS was supported in part by the MIT S3TEC Energy Research Frontier Center of the Department of Energy under Grant No. DE-SC0001299. This work was also supported in part by the Army Research Office through the Institute for Soldier Nanotechnologies under Contract No. W911NF-07-D0004.
\end{acknowledgments}

\newpage

\appendix

\section{{\large Supporting Information for: Enabling single-mode behavior over large areas with photonic Dirac cones.}}
\setcounter{equation}{0}
\setcounter{figure}{0}

In the following, we present a detailed description of the theoretical treatment of the spontaneous emission coupling efficiency (the $\beta$ factor) used in the main text. We start by deriving a general expression for the $\beta$ factor for the class of structures considered in our work. Within the Wigner-Weisskopf approximation~\cite{Scullybook}, the rate of spontaneous emission $\Gamma$  of a two-level quantum emitter embedded in a complex electromagnetic (EM) environment can be described by Fermi's \emph{golden rule} \cite{Loudonbook,Novotnybook}
\begin{equation}
\Gamma=\frac{\pi d^2 \omega_s}{3 \hbar \epsilon_0} \: \rho({\bf r}_s,\hat{{\bf d}},\omega_s) 
\end{equation}
where ${\bf d}=d \: \hat{{\bf d}}$ is the dipolar moment associated to the radiative transition of the emitter. Here $\omega_s$ is the frequency of that transition, whereas $\epsilon_0$ and $\hbar$ are the vacuum permittivity and the reduced Planck's constant, respectively. The scalar function ${\rho({\bf r},\hat{{\bf d}},\omega)}$ represents the photonic local density of states (LDOS) accessible to the emitter at position ${\bf r}={\bf r}_s$. 

On the other hand, in the case of a non-dissipative system, in which the electric field can be expanded in terms of a complete basis of transverse orthonormal modes $\{{\bf E}_{m}({\bf r})\}$ of frequencies $\{ \omega_m \}$, the LDOS can be expressed as
~\cite{Novotnybook,Busch98,Sprik96}
\begin{equation}
{\rho({\bf r},\hat{{\bf d}},\omega)}=\sum_{m} \delta(\omega-\omega_{m}) \:  \epsilon({\bf r})\: |{\hat {\bf d}} \: {\bf E}_{m}({\bf r})|^2
\end{equation}
where $\epsilon({\bf r})$ is the position dependent dielectric constant characterizing the system.
Each of the modes in Eq. (2) satisfies the following orthonormality condition
\begin{equation}
\int d{\bf r} \: \epsilon({\bf r}) \: {\bf E}_m({\bf r}) \: {\bf E}^*_n({\bf r})=\delta_{mn} 
\end{equation}
with $\delta_{nm}$ standing for the Kronecker's delta. The transversality condition reads
\begin{equation}
\nabla \cdot \left[ \epsilon({\bf r}) \: {\bf E}_m ({\bf r}) \right]=0
\end{equation}
Note also that each mode profile ${\bf E}_m ({\bf r})$ can be obtained by solving the following wave equation
\begin{equation}
\nabla \times \left[ \nabla \times {\bf E}_m({\bf r}) \right]=\mu_0 \epsilon({\bf r}) \omega_m^2 {\bf E}_m ({\bf r})
\end{equation}
where $\mu_0$ is the vacuum permeability.

Now, by definition, the $\beta$-factor can be calculated as $\beta=\Gamma_t/\Gamma_{all}$, where $\Gamma_t$ is the spontaneous emission rate into a given \emph{targeted} mode (often a laser mode) and $\Gamma_{all}$  is the total spontaneous emission rate into all the modes of the system (including the targeted one)\cite{VuckovicIEEE}. Thus, assuming that the lineshape of the emission transition is defined by function $g(\omega)$, by inserting Eq.~(2) into Eq.~(1) and integrating the resulting expression over $\omega$, we obtain the following expression for $\Gamma_{all}$, corresponding to an emitter located at ${\bf r}={\bf r}_s$
\begin{equation}
\Gamma_{all}=\frac{\pi d^2 \epsilon({\bf r}_s)}{3 \hbar \epsilon_0} \int d\omega \: g(\omega) \: \omega \: f(\omega,{\bf r}_s)
\end{equation}
where function  $f(\omega,{\bf r})$ is defined as
\begin{equation}
 f(\omega,{\bf r})=\sum_m \delta(\omega-\omega_m) \:  |{\hat {\bf d}} \cdot {\bf E}_{m}({\bf r})|^2
\end{equation}
Equations (6) and (7) summarize well the physical origin of the total spontaneous emission decay in the considered system: on the one hand, the different terms in the summand of Eq.~(7) account for the different modes in which a single frequency component $\omega$ of considered emission transition can decay to. On the other hand, the integral in $\omega$ appearing in Eq.~(6) accounts for the continuous sum of these possible radiative decay paths for all frequency components of the emission transition. Note that, as expected, the lineshape of the emission, $g(\omega)$, acts as a frequency dependent weight in this sum. The additional factor $\omega$ multiplying $g(\omega)$ in the integral of Eq.~(6) comes just from the proportionality factor that links the spontaneous emission rate and the LDOS (see Eq.~(1)).

This physical picture of the decay process also allows us to  obtain an expression for $\Gamma_t$ simply by singling out the contribution to $\Gamma_{all}$ that stems from the considered targeted mode. In particular, if we define ${\bf E}_t ({\bf r})$ and $\omega_t$ to be the targeted electric-field profile and its corresponding frequency, respectively, the magnitude of $\Gamma_t$ can be obtained by substituting $g(\omega)$ by $g_t(\omega) \equiv \delta(\omega-\omega_t)$ in Eq.~(7) and by replacing $f(\omega, {\bf r}_s)$ by $f_t(\omega, {\bf r}) \equiv \delta(\omega-\omega_t) |{\hat {\bf d}} \cdot {\bf E}_{t}({\bf r})|^2$. This yields
\begin{equation}
\Gamma_t=\frac{\pi d^2 \epsilon({\bf r}_s)}{3 \hbar \epsilon_0} \: \omega_t \: g(\omega_t) \:  |{\hat {\bf d}} \cdot {\bf E}_{t}({\bf r}_s)|^2
\end{equation}
Note that dividing Eq.~(6) by Eq.~(8), and using the definition of the LDOS given in Eq.~(2), we recover Eq.~(1) of the main text.

We now focus on the application of the above formalism to calculate the $\beta$ factor in the case in which the emitter is embedded in a three-dimensional photonic crystal (PhC). We assume that the considered PhC is characterized by a finite volume $V=L_x \times L_y \times L_z$ (where $L_x$, $L_y$, $L_z$ are the dimensions of the PhC along the $x$, $y$ and $z$ axis, respectively). Here we emphasize that in our theoretical calculations, this three-dimensional analysis is applied to the homogeneous and band-edge cases discussed in the main text. (The homogeneous case can be trivially considered as a periodic system with an arbitray periodicity). For the Dirac case, however, due to the out-of-plane subwavelength confinement of the EM fields introduced by the full photonic band gap, the analysis is performed in terms of the in-plane transverse area of the system, $A=L_x \times L_y$ (see discussion in the main text).
 
To analyze the finite-size effects on the $\beta$-factor, without loss of generality, we assume the volume $V$ (or transversal area $A$ for the Dirac case) to be surrounded by Born-von-Karman boundary conditions (i.e., periodic boundary conditions~\cite{kittelbook}; our theoretical analysis admits a straightforward generalization to other types of boundary conditions). In this case, the index $m$ used above to label the modes can be identified with $\{n,{\bf k},\sigma \}$, where $n$ is the band-index, ${\bf k}$ is the wavevector of each Bloch mode (${\bf k}$ lies inside the First Brillouin Zone (FBZ)) and $\sigma$ labels the polarization ($\sigma$=1 and $\sigma$=2, for $s$- and $p$-polarization, respectively). In addition, since the system is finite, ${\bf k}$ can only take discrete values: ${\bf k}=2 \pi \times (n_x/L_x, n_y/L_y, n_z/L_z)$ for the homogeneous and band-edge cases, and ${\bf k}=2 \pi \times (n_x/L_x, n_y/L_y, 0)$ for the Dirac case (in all three cases, $n_x$, $n_y$, and $n_z$ are arbitrary integers). Thus, once the normal modes of the system ${\bf E}_{n,{\bf k},\sigma}$ are computed (for which we have used the plane-wave expansion method to MaxwellÕs equations\cite{StevenOE}), from Eqs.~(6) and (7) the $\beta$ factor can be calculated from
\begin{equation}
\beta=\frac{\omega_t \: g(\omega_t)  \: |{\bf E}_t({\bf r}) \cdot {\hat{\bf d}}|^2} {\int d \omega \: g(\omega) \: \omega \: \left\{ \sum_{n,{\bf k},\sigma} \delta(\omega-\omega_{n,{\bf k},\sigma})\:  |{\bf E}_{n,{\bf k},\sigma}({\bf r}) \cdot {\hat{\bf d}}|^2 \right\} }
\end{equation}

In the limit in which the volume $V$ of the system is such that $V>>\lambda^3$ (or equivalently, for the Dirac case, when the area $A$ is such that $A>>\lambda^2$), with $\lambda$ being the central emission wavelength, semi-analytical expressions for the $\beta$-factor can be obtained by assuming a continuous distribution of ${\bf k}$ vectors over the FBZ. Specifically, we can replace $\sum_{\bf k} \rightarrow V/(2\pi)^3 \int_{FBZ} d{\bf k}$ in the homogeneous and band-edge cases, and $\sum_{\bf k} \rightarrow A/(2\pi)^2 \int_{FBZ} d{\bf k}$ in the Dirac case (note that in all three cases the integral over ${\bf k}$ is performed over the whole FBZ) . Then, if we expand the argument of the Dirac delta appearing in the denominator of the right-hand side of Eq. (9) using
\begin{equation}
\omega-\omega_{n,\sigma}({\bf k}_0)={\boldmath \nabla}_{{\bf k}_0} \omega [{\bf k}_0-{\bf k}_0(\omega_{n,\sigma})]+O(|{\bf k}_0-{\bf k}_0(\omega_{n,\sigma})|^2)
\end{equation}
and neglect the contribution of second order terms in $|{\bf k}_0-{\bf k}_0(\omega_{n,\sigma})|$~\cite{kittelbook}, after some straightforward algebra, one finds that Eq.~(9) can be rewritten as
\begin{equation}
 \beta=\frac{1}{V} \frac{\omega_s\:g(\omega_s)} {\int d\omega\:  g(\omega) \: \omega \: \tilde{\rho}(\omega)}  
\end{equation}
where the function $\tilde{\rho}(\omega)$ determines the total density of photonic states per unit volume in the structure. Note that in the Dirac case, $V$ must be replaced by the transversal area $A$. In analogy with standard analyses in solid-state physics~\cite{kittelbook}, in the homogeneous and band-edge cases, $\tilde{\rho}(\omega)$  can be expressed as
\begin{equation}
 \tilde{\rho}(\omega)=\frac{1}{(2\pi)^3}\: \int_{A(\omega_s)} \frac{1}{v_g} \: dk_t
\end{equation}
where $A(\omega_s)$ denotes the equifrequency surface $\omega=\omega_s$, and $v_g$ is the magnitude of the group velocity $v_g=|d\omega/dk|$. At each point of the equifrequency surface, $k_t$ stands for the component of the 3D vector ${\bf k}$ that lies along the tangential direction to $A(\omega_s)$ at each point of the $k$-space. In the Dirac case a similar expression holds for  $\tilde{\rho}(\omega)$, but now the domain of integration in Eq.~(12) is a equifrequency curve instead of equifrequency surface. The resulting expressions $\tilde{\rho}(\omega)$, obtained by performing the integral defined by Eq. (12) for the different dispersion relations considered in this work, are discussed in detail in the main text.

\begin{figure}[t]
\includegraphics[width=14cm]{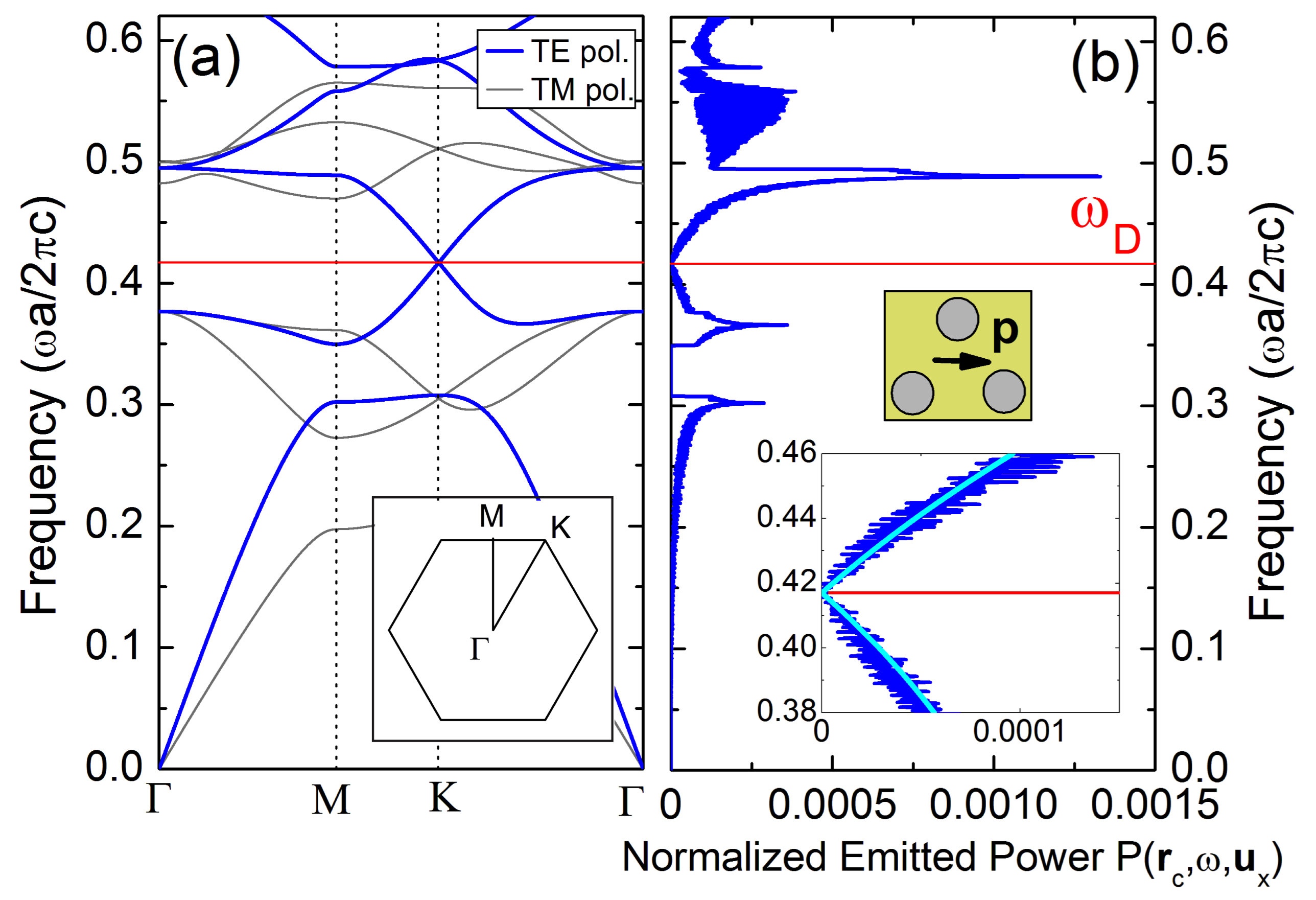}
\caption{(a) Numerical calculation of two-dimensional photonic bands displaying an isolated Dirac point. The analyzed system is formed by a triangular lattice of dielectric cylinders of refractive index $n_d=3.1$ and radius $r_d=0.32 a$ ($a$ being the lattice constant) embedded in air. (b) Power emitted by a dipole located at the center of the unit cell of the considered photonic crystal and with its dipolar moment pointing along the $x$-axis (see sketch in top inset of this panel). In both panels, $\omega_D$ marks the frequency of the Dirac point. Bottom inset in Fig.~1b shows the comparison between the predictions of semi-analytical and full numerical calculations (cyan and blue lines, respectively) for the emitted power.}
\end{figure}

Importantly, in deriving Eq.~(11) we have assumed that for the range of parameters considered in this work, the emission bandwidth is narrow enough so that $|{\bf E}_{n {\bf k} \sigma} ({\bf r})|^2 \approx |{\bf E}_t ({\bf r})|^2$ for all modes whose equifrequencies $\omega_{n{\bf k}\sigma}$ lie inside the interval where $g(\omega)$ is not negligible. To verify numerically the accuracy of this approximation in the Dirac case (a similar analysis holds for the band-edge case), we have probed directly the LDOS of the 2D counterpart of the defect layer structure shown in Fig.~1a of the main text. Specifically, in order to do that, we have computed the power radiated by a dipole placed in the low-refractive index regions of the structure (i.e., in the interstitial regions among cylinders), $P_{out}({\bf r},{\hat {\bf d}},\omega)$. To compute $P_{out}({\bf r},{\hat {\bf d}},\omega)$ we have employed a generalization of the conventional coupled-mode theory~\cite{Hausbook}, in which each Bloch mode is considered as an independent input/output channel (see details in Ref.~\cite{Rafif08}). The computed results are summarized in Fig.~1 of this Supplementary Information, in which band structure calculations (Fig.~1a) are displayed together with the corresponding dependence of $P_{out}({\bf r},{\hat {\bf d}},\omega)$  with frequency (Fig.~1b). As observed in these results, the dependence of $P_{out}({\bf r},{\hat {\bf d}},\omega)$ with $\omega$ near the Dirac frequency $\omega_D$ (and hence the LDOS~\cite{Yariv}), obtained by assuming $|{\bf E}_{n{\bf k}\sigma} ({\bf r})|^2 \approx |{\bf E}_s ({\bf r})|^2$, is in good agreement with full numerical calculations within a moderately large bandwidth of frequencies (see comparison between cyan line and blue line in the bottom inset of Fig.~1b). Finally, we note that although in the particular case considered in Fig.~1b the dipole has been placed at  the center of the unit cell of the triangular lattice, with ${\hat {\bf d}}$ pointing along the $x$-direction (see top inset of Fig.~1b), we have checked that similar good agreement between numerical and and semi-analytical results is obtained for other positions of the dipole in the unit cell, as well as for other orientations of ${\hat {\bf d}}$.



\begin{thebibliography}{99}

\bibitem{Novoselov04}
K. S. Novoselov, A. K. Geim, S. V. Morozov, D. Jiang, Y. Zhang, S. V. Dubonos, I. V. Grigorieva and A. A. Firsov, Science {\bf 306}, 666 (2004).

\bibitem{Zhang05}
Y. Zhang, Y. W. Tan, H. L. Stormer, and P. Kim,  Nature {\bf 438},  201 (2005).

\bibitem{Novoselov05}
K. S. Novoselov, A. K. Geim, S. V. Morozov, D. Jiang, M. I. Katsnelson, I. V. Grigorieva, S. V. Dubonos, and A. A. Firsov, Nature {\bf 438}, 197 (2005).

\bibitem{Berger06}
C. Berger, Z. Song, X. Li, X. Wu, N. Brown, C. Naud, D. Mayou, T. Li, J. Hass, A. N. Marchenkov, E. H. Conrad, P. N. Frist, W. A. de Heer, Science {\bf 312} , 1191 (2006).

\bibitem{Geim07}
A. K. Geim and K. S. Novoselov, Nat. Mat. {\bf 6}, 183 (2007).

\bibitem{CastroNeto09}
A. H. Castro Neto, F. Guinea, N. M. R. Peres, K. S. Novoselov, and A. K. Geim, Rev. Mod. Phys. {\bf 81}, 109 (2009).

\bibitem{Geim09}
A. K. Geim, Science {\bf 324}, 1530 (2009).

\bibitem{JJbook}
J. D.	Joannopoulos,	S. G.	Johnson,	J. N.	Winn, and R.D. Meade, 
\emph{Photonic Crystals: Molding the Flow of Light} (Princeton University Press, Princeton, NJ, 2008), 2nd ed.

\bibitem{Haldane08}
 F. D. M. Haldane and S. Raghu, Phys. Rev. Lett. {\bf 100}, 013904 (2008).
 
\bibitem{Sepkhanov07}
R. A. Sepkhanov, Ya. B. Bazaliy, and C. W. J. Beenakker, Phys. Rev. A {\bf 75}, 063813 (2007).

\bibitem{Dood10}
S. R. Zandbergen and M. J. A. de Dood, Phys. Rev. Lett. {\bf 104}, 043903 (2010).

\bibitem{Bahat-Treidel10}
O. Bahat-Treidel, O. Peleg, M. Grobman, N. Shapira, M. Segev, and T. Pereg-Barnea, Phys. Rev. Lett. {\bf 104}, 063901 (2010)

\bibitem{Zhang08}
X. Zhang, Phys. Rev. Lett. {\bf 100}, 113903 (2008)

\bibitem{Huang11}
X. Huang, Y. Lai, Z. H. Hang, H. Zheng, and C. T. Chan, Nat. Mat.  {\bf 10}, 582 (2011).

\bibitem{Johnson00}
S.G. Johnson and J.D. Joannopoulos, Appl. Phys. Lett. {\bf 77}, 3490 (2000).

\bibitem{Qi04}
Mi. Qi, E. Lidorikis, P. T. Rakich, S. G. Johnson, J. D. Joannopoulos, E. P. Ippen, and H. I. Smith, Nature {\bf 429}, 538 (2004).

\bibitem{Povinelli01}
M. L. Povinelli, S. G. Johnson, S. Fan, and J. D. Joannopoulos, Phys. Rev. B {\bf 64}, 075313 (2001).

\bibitem{Johnson01}
S. G. Johnson and J. D. Joannopoulos, Opt. Express {\bf 8}, 173 (2001).

\bibitem{Purcell}
E. Purcell, Phys. Rev. {\bf 69}, 37 (1946).

\bibitem{Novotnybook}
L. Novotny and B. Hecht, \emph{Principles of Nano-Optics} (Cambridge University Press, Cambridge, England, 2006).

\bibitem{Yariv}
Y. Xu, R. K. Lee, and A. Yariv, Phys. Rev. A {\bf 61},  033807 (2000).

\bibitem{Coldrenbook}
L. A. Coldren and S. W. Corzine, \emph{Diode lasers and photonic integrated circuits} (Wiley-Interscience, New York, 1995)

\bibitem{Yokoyama92}
H. Yokoyama, Science {\bf 256}, 66 (1992).

\bibitem{Baba91}
T. Baba, T. Hamano, and F. Koyama, IEEE J. Quantum Electron. {\bf 27}, 1347 (1991).

\bibitem{Painter99}
O. Painter, R. K. Lee, A. Scherer, A. Yariv, J. D. O'Brien, P. D. Dapkus, and I. Kim, Science {\bf 284}, 1819Ð1821 (1999).

\bibitem{Vuckovic06}
H. Altug, D. Englund, and J. Vuckovic, Nat. Phys. {\bf 2}, 484 (2006).

\bibitem{Vuckovic11}
B. Ellis, M. A. Mayer, G. Shambat, T. Sarmiento, J. Harris, E. E. Haller, and J. Vuckovic, Nat. Phot. {\bf 5}, 297 (2011).

\bibitem{Santori02}
C. Santori, D. Fattal, J. Vukovic, G. S. Solomon, and Y. Yamamoto, Nature {\bf 419}, 594 (2002).

\bibitem{Chang06}
W. H. Chang, W. Y.  Chen, H. S. Chang, T. P. Hsieh, J. I. Chyi, and T. M. Hsu, Phys. Rev. Lett. {\bf 96}, 117401 (2006).

\bibitem{kittelbook}
C. Kittel, \emph{Introduction to Solid State Physics} (Wiley, New York, 1976).

\end{thebibliography}

\begin{thebibliography}{99}

\bibitem{Scullybook} M. O. Scully and M. S. Zuibary, \emph{Quantum Optics} (Cambridge University Press, Cambridge, 1997).
\bibitem{Loudonbook} R. Loudon, \emph{The quantum theory of light} (Oxford Univ. Press, New York, 2000).
\bibitem{Novotnybook} L. Novotny and B. Hecht, \emph{Principles of Nano-Optics} (Cambridge University Press, Cambridge, England, 2006). 
\bibitem{Busch98} K. Busch and S. John, Phys. Rev. E {\bf 58} 3896 (1998).
\bibitem{Sprik96} R. Sprik, B. A. van Tiggelen, and A. Lagendijk, Europhys. Lett. {\bf 35}, 265 (1996).
\bibitem{kittelbook}
C. Kittel, \emph{Introduction to Solid State Physics} (Wiley, New York, 1976).
\bibitem{VuckovicIEEE}
J. Vuckovic, O. Painter, Y. Xu, and A. Yariv, IEEE J. of Quantum Electr. {\bf 35},  1168 (1999). 
\bibitem{StevenOE}
S. G. Johnson and J. D. Joannopoulos, Opt. Express {\bf 8}, 173 (2001).
\bibitem{Hausbook}
H. A. Haus, \emph{Waves and Fields in Optoelectronics} (Prentice-Hall, Englewood Cliffs, NJ, 1984).
\bibitem{Rafif08}
R. Hamam, M. Ibanescu, E.J. Reed, P. Bermel, S.G. Johnson, E. Ippen, J. D. Joannopoulos, and M. Soljacic, Opt. Express {\bf 16}, 12523 (2008).
\bibitem{Yariv}
Y. Xu, R. K. Lee, and A. Yariv, Phys. Rev. A {\bf 61},  033807 (2000).
\end{thebibliography}
\end{document}